\documentclass[12pt]{iopart}
\usepackage{iopams}  

\newcommand\ro{\hat\rho}
\newcommand\Oo{\hat O}
\newcommand\rO{\widehat\rho}
\newcommand\Mo{\hat M}
\newcommand\Po{\hat P}
\newcommand\Ho{\hat H}
\newcommand\HO{\widehat H}
\newcommand\Lo{\hat L}
\newcommand\LO{\widehat L}
\newcommand\dg{^\dagger}
\newcommand\half{{\frac{1}{2}}}

\newcommand\qo{\hat q}

\newcommand\der{\partial}
\newcommand\Io{\hat I}
\newcommand\ket[1]{\left|#1\right>}
\newcommand\bra[1]{\left<#1\right|}

\newcommand\expect[1]{\left<#1\right>}

\begin{document}

\title{Hybrid Quantum-Classical Master Equations}

\author{Lajos Di\'osi}

\address{Wigner Research Center for Physics,
H-1525 Budapest 114, POB 49, Hungary}
\ead{diosi.lajos@wigner.mta.hu}

\begin{abstract}
We discuss hybrid master equations of composite systems which are hybrids of classical 
and quantum subsystems. A fairly general form of hybrid master equations is suggested, 
its consistency is derived from the consistency of Lindblad quantum master equations. 
We emphasize that quantum measurement is a natural example of exact hybrid systems.
We derive a heuristic hybrid master equation of time-continuous position measurement 
(monitoring).
\end{abstract}
%\maketitle

\section{Introduction}
\label{Introduction}
Notion of hybrid systems---consisting of quantum \emph{and} classical subsystems---arises in different contexts, see, e.g., refs. 
\cite{Imretal67,SheSud78,Ale81,Bou88,Dio95,Sal96,Dio96,PreKis97,KapCic99,Dio00b,Dio00c,DioGisStr00,PerTer01,Kap06,Dio08,Tsa09,Dio11,Elz12}.
We are interested in those works which are using the notion of hybrid density explicitely or imlicitely at least, 
with a standard statistical interpretation.
Hybrid dynamics in the narrow sense means dynamical coupling between a quantum and a classical dynamical system,
sometimes the latter one is just another quantum subsystem described in classical phase space variables, but in
all versions the coupling results in irreversibilities \cite{Imretal67,Ale81,Dio96,KapCic99,DioGisStr00}.
Hybrid formalism of quantum measurement means the irreversible interaction between
a quantum dynamical system and the classical pointer of the measuring device, cf., e.g., \cite{SheSud78,DioGisStr00,Dio08,Tsa09}.
Hybrid dynamics in the general sense means \emph{any}---not necessarily canonical, unitary, or even reversible---dynamics of coexisting quantum and
classical states and variables, cf. \cite{Dio00b,Dio00c}. 

The mathematical representation of hybrid systems unifies 
the mathematical representations of classical and quantum
systems, respectively. The notion of hybrid density $\ro(x)$, to represent
the hybrid state, follows from the notions of classical density $\rho(x)$ and
quantum density matrix $\ro$ in a straightforward way, see section \ref{Hybrid_density} for
the rigorous definition. An exact application of hybrid formalism is no doubt the action
of quantum measurement (section \ref{Measurement}). The hybrid dynamical equation (master equation, ME) is
an open issue. One can profit from the generic Pauli and Lindblad  MEs of separate classical and quantum systems, respectively.
We shall offer a partial solution as to the general structure of hybrid MEs, see sections \ref{Hybrid_dynamics} and \ref{Derivation}. 
The lessons are applied in section \ref{Quantum_monitoring} to construct the hybrid ME of time-continuous measurement (monitoring).
\begin{table}[h]
\caption{\label{tab1}States and dynamics in classical, quantum, and hybrid systems.} 
\begin{indented}
\lineup
\item[]\begin{tabular}{l c c c}
\br                              
&Classical&Quantum&Hybrid\\ 
\mr
Density:   &$\rho(x)$&$\ro$&$\ro(x)$\\
Master Eq.:&Pauli&Lindblad&?\cr 
\br
\end{tabular}
\end{indented}
\end{table}

\section{Hybrid density}
\label{Hybrid_density}
As we mentioned, the general hybrid system consists of a quantum system
and of any,  not necessarily dynamical classical system, including discrete as
well as continuous classical systems.  Let us, e.g., consider a classical
system described by a discrete variable $x$ of probability density $\rho(x)$,
together with an independent quantum system of state $\ro$. 
To model their coexistence, we form their hybrid system whose
hybrid state must be
\begin{equation}\label{HDprod}
\ro(x)=\rho(x)\ro\;.
\end{equation}
The general, correlated, hybrid state must be positive semidefinite:
\begin{equation}\label{HD}
\ro(x)\geq0,~~\forall x
\end{equation}
and normalized:
\begin{equation}\label{HDnorm}
\Tr\sum_x\ro(x)=1\;.  
\end{equation}
The conditions (\ref{HD},\ref{HDnorm}) are necessary and sufficient  for $\ro(x)$ 
to be a legitimate hybrid density.

We define the reduced state (density matrix) of the quantum subsystem by
\begin{equation}\label{QDred}
\ro=\sum_x\ro(x)\;,
\end{equation}
the reduced state (density) of the classical subsystem by
\begin{equation}\label{CDred} 
\rho(x)=\Tr\ro(x)\;,
\end{equation}
and the conditional state (density matrix) of the quantum subsystem by
\begin{equation}\label{QDcond}
\ro_x=\ro(x)/\rho(x)\;.
\end{equation}
Note that the conditional state of the classical subsystem is pointless
since the quantum subsystem does not feature conditions unless we
perform a quantum measurement on it (table \ref{tab1}).
\begin{table}[h]
\caption{\label{tab2}Reduced classical and quantum states. Conditional classical state. 
Missing conditional quantum state.}
\begin{indented}
\lineup
\item[]\begin{tabular}{c c c c}
\br                              
Reduced Q              &Reduced C              &Conditional C              &Conditional Q\\ 
\mr
$\ro\!=\!\sum_x\ro(x)$ &$\rho(x)\!=\!\Tr\ro(x)$ &$\ro_x\!=\!\ro(x)/\rho(x)$ &$\not\exists$\\
\br
\end{tabular}
\end{indented}
\end{table}

The statistical interpretation of the hybrid density is straightforward from
the statistical interpretation of density matrices and classical densities.
Let $\Oo(x)=[\Oo(x)]\dg$ stand for the generic hybrid observable,
its expectation value can be calculated as follows, cf. e.g., \cite{Dio96}:
\begin{equation}\label{Hstat}
\expect{\Oo(.)}_{\ro(.)}=\Tr\sum_x \Oo(x)\ro(x)\;.
\end{equation}

Having outlined the abstract features, let's see an occasional list of very different hybrid systems  
with their respective hybrid densities. In molecular physics $\ro$ refers to quantized electrons and 
$(r,p)$ stands for classical nuclear positions and momenta yielding $\ro(r,p)$ \cite{Ale81}. 
In quantum optics  $\ro$ refers to quantized electrons and $a,a^\star$ stand for the complex amplitudes 
of classical e.m. field modes yielding $\ro(a,a^\star)$ \cite{Dio96}. In nanophysics  $\ro$ refers 
to the quantum dot and  $n$ stands for the charge count yielding $\ro(n)$. Last but not least, 
in quantum measurement $\ro$ refers to the measured quantum system and $x$ stands for the measurement 
outcome yielding $\ro(x)$ \cite{DioGisStr00}, as discussed below.

\section{Measurement}
\label{Measurement}
What happens to the quantum state $\ro$ of a quantum system, under measurement of the complete set of orthogonal projectors $\{\Po_x\}$? 
We argue that hybrid formalism and interpretation are exact alternatives to the standard ones in textbooks. 
Textbook formalism says that the pre-measurement state $\ro$ jumps randomly to the post-measurement conditional quantum state $\ro_x$: 
\begin{equation}\label{vN}
\ro\longrightarrow \ro_x=\frac{1}{\rho(x)}\Po_x\ro\Po_x
\end{equation}
with probability $\rho(x)=\Tr(\Po_x\ro\Po_x)$. 

In hybrid formalism, we say that $\ro$ jumps deterministically  into the post-measurement hybrid state:  
\begin{equation}\label{HvN}
\ro\longrightarrow\ro(x)=\Po_x\ro\Po_x\;.
\end{equation}
The randomness of the outcome $x$ is now expressed through the 
statistical interpretation (\ref{Hstat}), as well as (\ref{QDred}-\ref{QDcond}), of the hybrid state.

In complete generality, hybrid formalism is convenient for general (unsharp)
quantum measurements defined by Kraus operators $\Mo_x$ instead of
projectors $\Po_x$, satisfying completeness $\sum_x \Mo_x\dg\Mo_x=\Io$
but no orthogonality or hermiticity. In hybrid formalism,  measurement is 
fully represented by the jump of the pre-measurement  quantum state  $\ro$ into 
the post-measurement hybrid state:
\begin{equation}\label{HKra}
\ro\longrightarrow\ro(x)=\Mo_x\ro\Mo_x\dg\;.
\end{equation}
The statistical interpretation of $\ro(x)$ reproduces the common rules (\ref{vN}) of
measurement.

A remarkable example is the Gaussian unsharp position measurement
whose Kraus operators are
\begin{equation}\label{Krapos}
\Mo_x=\Mo_x\dg=(2\pi\sigma^2)^{-1/4}\exp\left[-\frac{(\qo-x)^2}{4\sigma^2}\right]\;,
\end{equation}
i.e., the square roots of unsharp 'projectors' labeled by their central positions $x$
(which are continuous classical variables this time). 
Their measurement in hybrid formalism (\ref{HKra}) reads:
\begin{equation}\label{HKrapos}
\ro\longrightarrow\ro(x)
=\frac{1}{\sqrt{2\pi\sigma^2}}\exp\left[-\frac{(\qo-x)^2}{4\sigma^2}\right]\ro\exp\left[-\frac{(\qo-x)^2}{4\sigma^2}\right]\;.
\end{equation}
This is the key to the time-continuous position measurement (monitoring) theory \cite{Dio88,BarBel91}, 
for a heuristic derivation of the corresponding hybrid ME see section \ref{Quantum_monitoring}.   
We study general hybrid MEs first.

\section{Hybrid dynamics}
\label{Hybrid_dynamics}
All Markovian classical MEs must have the Pauli form \cite{Pau64}:
\begin{equation}\label{CME}
\frac{d\rho(x)}{dt}=\sum_{y}[T(x,y)\rho(y)\!-\!T(y,x)\rho(x)]\;,
\end{equation}
where $T(x,y)\geq0$ is an arbitrarily given transition rate from $y$ to $x$.
Note for completeness that also a drift term $-\der_x v(x)\rho(x)$ of 
arbitrarily given drift velocity $v(x)$ can be added to the r.h.s. when $x$ is continuous variable.  
In this case, the transition rates $T(x,y)$ can be
smooth non-negative functions, but the particularly important diffusion
process requires the singular ones:
\begin{equation}\label{Difftrans}
T(x,y)=\lim_{\tau\rightarrow0}\frac{1/\tau}{\sqrt{4\pi D\tau}}\exp\left[-\frac{(x-y)^2}{4D\tau}\right]\;.
\end{equation}
Substituting this form into the ME (\ref{CME}) yields the standard diffusion ME:
\begin{equation}\label{DifME}
\frac{d\rho}{dt} =D\der_x^2\rho(x)\;
\end{equation} 
with the diffusion coefficient $D$.

All Markovian quantum MEs must have the Lindblad form \cite{Lin76}:
\begin{equation}\label{QME}
\frac{d\ro}{dt}=-i[\Ho,\ro]+\sum_{\alpha}[\Lo_{\alpha}\ro\Lo_{\alpha}\dg-\half\{\Lo_{\alpha}\dg\Lo_{\alpha},\ro\}]\;,
\end{equation}
where  $H$ is the Hamiltonian, $\Lo_\alpha$ are arbitrarily given Lindblad operators (transition amplitudes). 
In a particular simple case, we have a single Hermitian Lindblad operator proportional to the position operator $\qo$ 
of a particle: $\Lo=\Lo\dg=\sqrt{2D'}\qo$. This yields
\begin{equation}\label{DecME}
\frac{d\ro}{dt}=-i[\Ho,\ro]-D'[\qo,[\qo,\ro]]\;,
\end{equation}
which describes momentum diffusion and, equivalently, position decoherence with coefficient $D'$. Note this quantum ME
governs the particle's quantum state under time-continuous unsharp position measurement, i.e., when 
position measurements (\ref{HKrapos}) of infinite unsharpness $\sigma^2\rightarrow\infty$ are repeated at infinite 
frequency $\nu\rightarrow\infty$ while $\nu/8\sigma^2=D'$ is kept fixed, cf., e.g., in \cite{Dio88}.   

The most generic form of the Markovian hybrid `Pauli-Lindblad' ME is not known, for particular results 
in very different contexts see, e.g., refs. \cite{Dio95,DioGisStr00,Kap06} among many others.
We guess a large class can be of the following structure:
\begin{eqnarray}\label{HME} 
\frac{d\ro(x)}{dt}=\!&-&\!\!i[\Ho(x),\ro(x)]\nonumber\\
                  \!&+&\!\!\sum_{y,\alpha}\left[\Lo_\alpha(x,y)\ro(y)\Lo_\alpha\dg(x,y)-\half\{\Lo_\alpha\dg(y,x)\Lo_\alpha(y,x),\ro(x)\}\right]
\end{eqnarray}
with completely arbitrary hybrid Pauli-Lindblad transition amplitudes $\Lo_\alpha(x,y)$. 

We learned before that the classical system can be discrete or continuous, and in the latter case 
the transition rates $T(x,y)$ can be smooth or singular as well. Similar features can occur 
to  the hybrid transition amplitudes $L_\alpha(x,y)$. For a $\delta'(x-y)$ singularity, 
we shall consider a particular example, quantum position monitoring, in section \ref{Quantum_monitoring}. 
In the forthcoming section, however,  we prove the consistency of (\ref{HME}) in the special case of discrete functions $\Lo_\alpha(x,y)$. 

\section{Derivation of the hybrid master equation}
\label{Derivation}
We are going to embed the hybrid ME (\ref{HME}) into the Lindblad ME (\ref{QME}) of a bigger quantum system by formal 
re-quantization of the classical subsystem. To this end, the Hilbert space
spanned by the basis vectors $\ket{x}$ is introduced. Then we upgrade
the hybrid state, Hamiltonian, and transition generators into composite
operators on the big Hilbert space:
\begin{eqnarray}
\ro(x)\rightarrow{\rO}&=&\sum_x\ro(x)\otimes\ket{x}\bra{x}\label{HDup}\\
\Ho(x)\rightarrow{\HO}&=&\sum_x\ro(x)\otimes\ket{x}\bra{x}\label{HHup}\\
\Lo_\alpha(x,y)\rightarrow\LO_\alpha&=&\sum_{x,y}\Lo_\alpha(x,y)\otimes\ket{x}\bra{y}\label{HLup}\;.
\end{eqnarray}
Now we consider the following Lindblad ME:
\begin{equation}\label{QMEbig}
\frac{d\rO}{dt}=-i[\HO,\rO]+\sum_\alpha\left[\LO_\alpha\rO\LO_\alpha\dg-\half\{\LO_\alpha\dg\LO_\alpha,\rO\}\right]\;.
\end{equation}
It is consistent, as we know. By construction, it preserves the 
block diagonality (\ref{HDup}) of $\rO$. So, if we multiply both sides by $\Io\otimes\ket{x}\bra{x}\dots$ 
and take the partial trace on both sides, we get exactly the hybrid ME (\ref{HME}). Therefore the consistency 
of the latter is guaranteed by the Lindblad ME (\ref{QMEbig}). This a central result of our work.

\section{Quantum monitoring}
\label{Quantum_monitoring}
We are going to construct the fenomenological hybrid equations of quantum monitoring.
Suppose we are continuously measuring (monitoring) the position $\qo$.
The classical variable $X$ will encode the monitored value so we introduce
the hybrid density $\ro(X)$ to represent the joint statistics of the particle
quantum observables and the monitored value of $\qo$. We are looking
for the dynamics of $\ro(X)$.

The hybrid ME that evolves $\ro(X)$ should contain a coupling between
 $\qo$ and $X$. Heuristically, we take the following naive ME:
\begin{equation}\label{HMEmonnaive}
\frac{d\ro(X)}{dt}=-i[\Ho,\ro(X)]-\half\der_X\{\qo,\ro(X)\}\;.
\end{equation}
The coupling term yields the following relationship:
\begin{equation}\label{Xq}
\frac{d\expect{X}}{dt}=\expect{\qo}\;.
\end{equation}
So far so good: the statistics of $X$ provides relevant and transparent information on the  
potential values of the position $\qo$. Our heuristic model seems to work.

There is a problem, however. The structure  $-\half\der_X\{\qo,\ro(X)\}$ is known to violate 
the  positivity of $\ro(X)$ \cite{DioGisStr00}. Our naive hybrid ME
is not correct. Nonetheless, we can find the correct one. Invoking the method
 of section \ref{Derivation}, we postulate the Lindblad ME (\ref{QMEbig}) on the big Hilbert space, 
with a carefully choice of a single Lindblad operator:
\begin{equation}
\LO=\qo/\sqrt{8D}\otimes\Io+\sqrt{2D}\Io\otimes\int\!\!(\der_X\!\ket{X})\!\bra{X}dX\;.
\end{equation}
Substitute this into (\ref{QMEbig}), multiply both sides by $\Io\otimes\ket{X}\bra{X}$ and take partial
trace on both sides, you get a correct hybrid ME of position monitoring:
\begin{eqnarray}\label{HMEmon}
\frac{d\ro(X)}{dt}=\!&-&\!\!i[\Ho,\ro(X)]-\half\der_X\{\qo,\ro(X)\}\\
                   \!&+&\!\!D\der_X^2\ro(X)-\frac{1}{16D}[\qo,[\qo,\ro(X)]]\;.\nonumber
\end{eqnarray}
Note the appearance of two additional diffusion terms on the r.h.s. which cure the defect of the naive ME (\ref{HMEmonnaive}). 
The crucial relationship (\ref{Xq}) between the classical variable $X$ and the monitored  position $\qo$
survives whereas a diffusive noise is superposed on the measured signal $X$ as well as on the momentum of the particle.
Observe the exact reciprocal trade between the signal noise and momentum diffusion (i.e.: position decoherence). 

Let's integrate both sides over $X$. We find that the reduced quantum state $\ro$ obeys the simple Lindblad ME (\ref{DecME}) 
with $D'=1/16D$. A more complex derivative of the hybrid ME of monitoring is the following non-autonomous Fokker-Planck equation 
for the reduced classical density:
\begin{equation}\label{Xqcorr}
\frac{d\rho(X)}{dt}=-\der_X\!\expect{\qo}_X\!\rho(X)\!+\!D\der_X^2\rho(X)\;,
\end{equation}
where $\expect{\qo}_X=\Tr(\qo\ro_X)$ is the conditional expectation value of $\qo$. This equation expresses
the diffusive noise superposed on the measured signal $X$, at diffusion coefficient $D$.

The hybrid ME (\ref{HMEmon}) is equivalent with the standard theory of quantum monitoring \cite{Dio88}
which prescribes two coupled Ito stochastic differential equations for the conditional state $\ro_X$
and the measured signal $X$, respectively, instead of the hybrid ME for $\ro(X)$. The proof of equivalence of
the two Ito equations with the hybrid ME is straightforward, will be shown elsewhere.

\section{Summary, outlook}
We gave a short introduction into the concept of hybrid systems.
We emphasized  that quantum measurement has a natural hybrid
formalism. A novel general structure of hybrid ME has been proposed, to unify the
Pauli and Lindblad structures. An application to quantum continuous 
measurement (monitoring) has been shown.

Ad hoc hybrid theories are often falling short.
Certain ad hoc MEs violate the positivity of the hybrid density. This failure is
abandoned by our class of hybrid MEs. Certain ad hoc hybrid theories don't
pass the "free will test" \cite{Dio12}. The measurement-related theories,
where the classical variable is the measured outcome, do pass it.
Nonetheless, the consistency of hybrid theories is being under discussion,
there can be a number of further consistency tests \cite{Elz13}.
 
The present work is a deliberate outline of certain important features of 
hybrid systems, with an emphasis on quantum measurement. Some novel results,
including the general structure of hybrid ME and the hybrid ME of quantum monitoring,
will be detailed and further clarified in forthcoming works \cite{Pre14} 

\ack
Support by the Hungarian Scientific Research Fund under Grant No. 75129 and support by the
EU COST Action MP1006 are acknowledged. The author is grateful to
Thomas Konrad and Sandeep Goyal for many useful discussions.

\end{document}